\documentstyle[11pt,newpasp,twoside,epsf]{article}
\def\ast{\mathchar"2203} \mathcode`*="002A  
\def\ltwig{\mathrel{\spose{\lower 3pt\hbox{$\mathchar"218$}} 
     \raise 2.0pt\hbox{$\mathchar"13C$}}} 
\def\gtwig{\mathrel{\spose{\lower 3pt\hbox{$\mathchar"218$}}  
     \raise 2.0pt\hbox{$\mathchar"13E$}}} 
\def\spose#1{\hbox to 0pt{#1\hss}} 
\reversemarginpar

\begin{document}
    
\vspace*{1cm}
\title{Magnetic Spin-Up of Line-Driven Stellar Winds} 
\author{Stan Owocki$^{1,2}$ and Asif ud-Doula$^{3}$
}
\affil{$^1$
Bartol Research Institute, University of Delaware, Newark, DE 
19350 USA
}
\affil{$^2$
Department of Physics and Astronomy, University of 
Glasgow, Glasgow G12 8QQ UK
}
\affil{$^3$
Department of Physics, North Carolina State University, 
Raleigh, NC 27695-8202 USA
}

\begin{abstract}
We summarize recent 2D MHD simulations of line-driven stellar winds from 
rotating hot-stars with a dipole magnetic field  aligned to the 
star's rotation axis.
For moderate to strong fields, much wind outflow is initially along
closed magnetic loops that nearly corotate as a solid body with the
underlying star, thus providing a torque that results in an
effective angular momentum spin-up of the outflowing material.
But instead of forming the ``magnetically torqued disk'' (MTD) postulated 
in previous phenemenological analyses, the dynamical simulations here 
show that material trapped near the tops of  such closed loops tends either 
to fall back or break out, depending on whether it is below or above the 
Keplerian corotation radius.
Overall the results raise serious questions about whether magnetic torquing 
of a wind outflow could naturally result in a Keplerian circumstellar disk.
However, for very strong fields, it does still seem possible to form a 
centrifugally supported, ``magnetically rigid disk'' (MRD),
in which the field not only forces material to maintain a rigid-body rotation, 
but for some extended period also holds it down against the outward 
centrifugal force at the loop tops.
We argue that such rigid-body disks seem ill-suited to explain the 
disk emission from Be stars, but could provide a quite attractive 
paradigm for circumstellar emission from the magnetically strong Bp 
and Ap stars.

\end{abstract}

\section{Background and Context}

Within the broad theme of this conference on ``Magnetic Fields in O, B, 
and A Stars'', a key issue is the role of magnetic fields in 
channeling the line-driven stellar wind outflows from such 
hot, luminous stars.
Initial results of full MHD simulations for models without stellar 
rotation  (ud-Doula and Owocki 2002) are reviewed in these proceedings 
through the write-up for the talk presented by A. ud-Doula.
Here we summarize further simulation results for {\it rotating} hot-stars with a 
magnetic dipole aligned to the stellar rotation axis.
These simulations provide a timely dynamical test of the ``Magnetically 
Torqued Disk'' (MTD) paradigm, which, as reviewed in the talk by J. Brown in 
these proceedings, has recently been promoted as a central mechanism for 
producing the disks inferred in Be stars (Cassinelli et al. 2002).

To allow greater emphasis on our MHD results, the broader discussion
of alternative Be-disk models given in the  oral version of the present paper 
will not  be reproduced here, as most of this has already been included in the 
write-up for a related talk presented at the recent IAU Symposium 215 on 
{\it Stellar Rotation} (Owocki 2003).
A key general result of these MHD simulations regards the tendency for 
magnetic torquing of wind material to result in centrifugal mass 
ejection rather a circumstellar disk.

\section{ Alfven Radius vs. Keplerian Rotation Radius}

Recent MHD  simulations (ud-Doula 2002; ud-Doula and Owocki 2002) indicate that 
the effectiveness of magnetic fields in channeling a stellar wind outflow 
can be characterized by the ratio of the magnetic to wind energy densities
\begin{equation}
    \eta (r) \equiv { B^2/8 \pi \over \rho v^{2} /2 } 
    = \eta_{\ast} 
    { (r/R_{\ast} ) ^{2-2q} \over (1-R_{\ast}/r)^{\beta} } \, .
\label{etadef}
\end{equation}
Here 
$\eta_{\ast} \equiv  B_{\ast}^{2}  R_{\ast}^{2} /( {\dot M } 
v_{\infty} )$ defines an overall ``magnetic confinement parameter'' 
in terms of the strength of the equatorial field $B_{\ast}$ at the stellar 
surface radius $R_{\ast}$, and the wind terminal momentum 
${\dot M} v_{\infty}$.
The latter equality thus isolates the radial variation in terms of a 
magnetic power-law index $q$ ($=3$ for a dipole) and a 
velocity index $\beta$ ($ \approx 1$ for a standard CAK wind).
If, for simplicity, we ignore the wind velocity variation (i.e. by 
taking $\beta=0$), we can easily solve for an ``Alfven radius'' 
$\eta(R_{A}) \equiv 1$ at which the magnetic and wind energy densities 
are equal
\begin{equation}
    R_{A} = \eta_{\ast}^{1/4} R_{\ast} \, .
\label{radef}
\end{equation}
As shown by simulation results summarized below, this Alfven radius
provides a reasonable estimate for the maximum radius of closed
loops in a wind outflow.
Moreover, since in rotating models such closed loops tend to keep the 
outflow in rigid-body rotation with the underlying star, it also 
defines the radius of maximum rotational spin-up of the wind azimuthal 
speed.

To characterize such rotational effects, let us next define a Keplerian
corotation radius $R_{K}$ at which rigid-body rotation would yield an 
equatorial centrifugal acceleration that just balances the
local gravitational acceleration from the underlying star,
\begin{equation}
  { G M \over R_{K}^{2} } = 
 { v_{\phi}^{2} \over R_{K} } =
 { V_{eq}^{2} R_{K} \over R_{\ast}^{2} } \, ,
\label{rkeqn}
\end{equation}
where
$V_{eq}$ is the stellar surface rotation speed at the equator.
This can be solved to yield
\begin{equation}
    R_{K} = W^{-2/3} R_{\ast} ,
\label{rkdef}
\end{equation}
where $W \equiv V_{eq}/V_{crit}$, 
with
$V_{crit} \equiv \sqrt{GM/R_{\ast}}$ the critical rotation speed.

Finally, it is also worth noting here that corotation out to an only slightly 
higher ``escape radius'',
\begin{equation}
    R_{E} = 2^{1/3} R_{K}  = 2^{1/3} W^{-2/3} R_{\ast} \, ,
\label{redef}
\end{equation}
would imply an azimuthal speed that equals the local escape speed from 
the star's gravitational field. 

\section{MHD Models with Parameters Optimized for Keplerian Spin-up}

The above scalings suggest that a likely necessary condition for propelling 
outflowing material into a Keplerian disk is to choose a combination 
of parameters for magnetic confinement vs. stellar rotation such that 
$R_{K} < R_{A} < R_{E}$.
In the parameter plane of $\sqrt{\eta_{\ast}} \sim B_{\ast}$ vs.  $W$ 
defined in figure 1, the required combination is represented by the 
relatively narrow gray band.
The dark region below this represents parameter combinations for 
which the azimuthal speed at the Alfven radius is expected to be 
sub-Keplerian, while the white region above represents cases for 
which the rotation speed at the Alfven radius should exceed the 
local gravitational escape speed.

As a sample test case, we focus here on the specific combination 
$\eta_{\ast}=10$ and $W=1/2$, which as shown by the dot in figure 1, 
lies in the middle of the gray domain, and thus should represent an 
optimal case for magnetic spin-up into Keplerian orbit. 
\begin{figure}
\begin{center}
\mbox{\epsfxsize=0.6\textwidth\epsfysize=0.25\textwidth\epsfbox{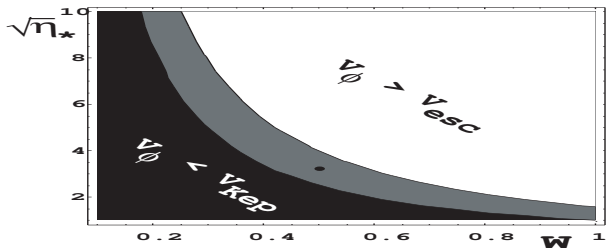}}
\caption{
The key domains in a parameter plane of magnetic field strength (as represented by 
$\sqrt{\eta_{\ast}} \sim B_{\ast}$) vs. stellar rotation (as represented by the critical 
rotation fraction $W$).
}
\vspace{-0.2cm}
\end{center}
\end{figure}
\begin{figure}
\begin{center}
\mbox{\epsfxsize=.7\textwidth\epsfysize=.7\textwidth\epsfbox{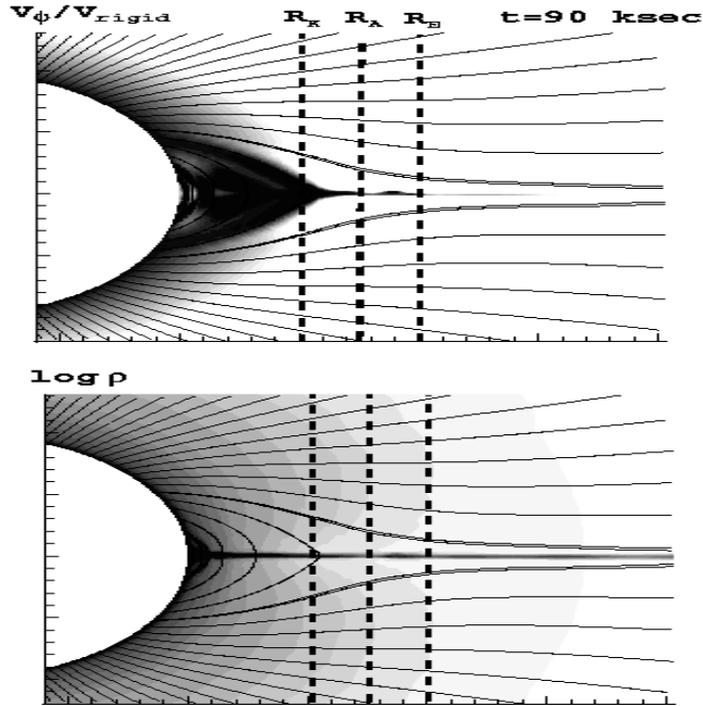}}
\caption{
Results of a 2D MHD simulation for a model with $\eta_{\ast} = 10$ and $W=1/2$, 
shown at at time snapshot 90 ksec after a dipole field is introduced into an 
initially steady-state, unmagnetized, line-driven wind.
In the lowel panel the grayscale represents the log of the mass density, 
while in the upper panel this represents the ratio of the azimuthal speed 
$V_{\phi}$ to the local rigid-body rotation speed,
$V_{rigid} = V_{rot} \sin \theta \, (r/R_{\ast} )$, with white to 
black corresponing to the range 0.7 to 1.
The curves denote magnetic field lines, and the vertical 
dashed lines indicate the equatorial location of the Keplerian, Alfven, and 
Escape radii defined in eqns. (2), (4), and (5) of the text.
}
\vspace{-0.2cm}
\end{center}
\end{figure}

Figures 2 and 3 illustrate results of 2D MHD simulations for this case, 
using the approach and general model assumptions described in ud-Doula 
and Owocki (2002), but now extended  to include field-aligned rotation.
Figure 2 shows that conditions at a time 90 ksec after 
introduction of the field do superficially resemble a magnetically 
torqued disk.
In the upper panel the darkest regions -- which represent a near rigid-body 
rotation with  $V_{\phi}/V_{rigid} \approx 1$ -- include essentially 
the entire closed magnetic loop.
The log-density grayscale in the lower panel shows moreover that magnetic 
channeling of the outflowing  stellar wind leads to formation of a dense 
equatorial compression that might be interpreted as the expected circumstellar 
disk.

\begin{figure}
\begin{center}
\mbox{\epsfxsize=1\textwidth\epsfysize=1\textwidth\epsfbox{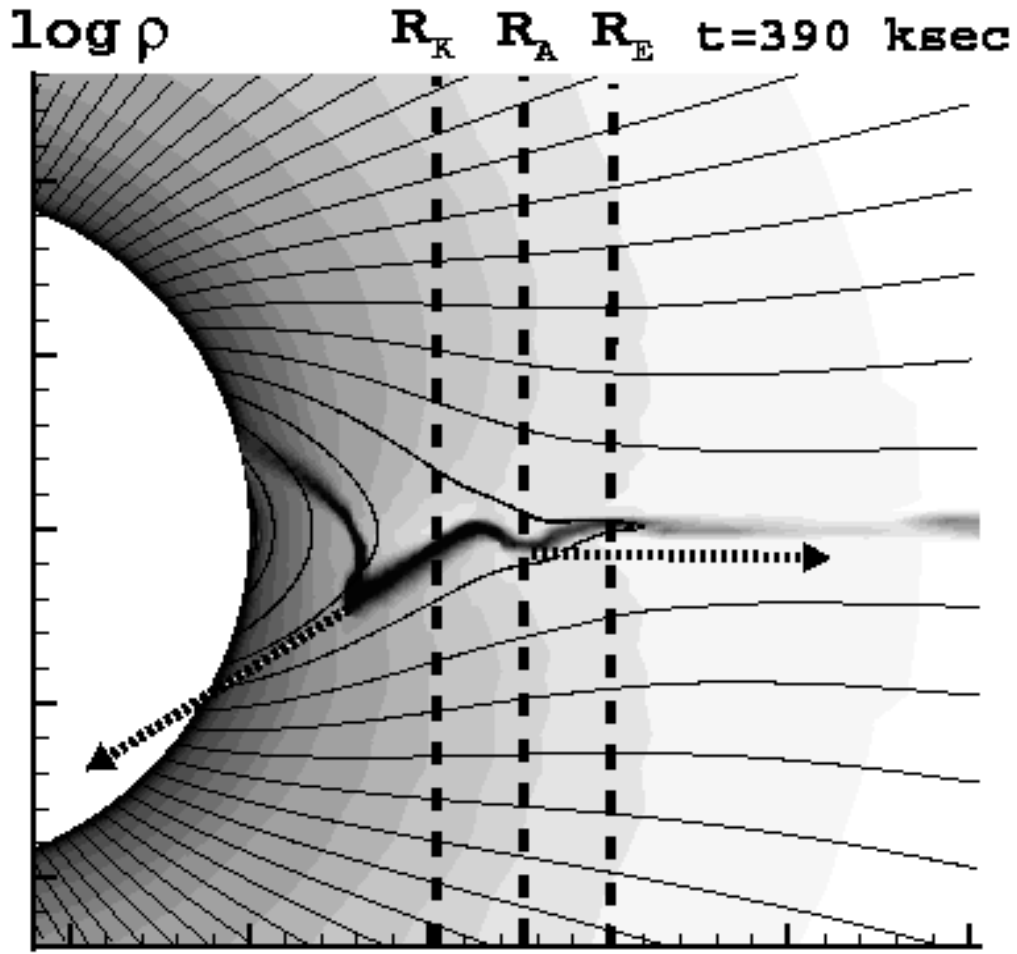}}
\caption{
Same as lower panel of figure 2, except for time t=390 ksec, with 
arrows illustrating the upward and downward flow direction of dense
material above and below the Keplerian radius.
}
\vspace{-0.2cm}
\end{center}
\end{figure}

Closer examination shows, however, that  most of this equatorial compression 
does not have the appropriate velocity for a stable, stationary, Keplerian orbit.
Thus in just a few ksec of subsequent evolution, this putative 
``disk'' becomes completely disrupted, characterized generally by 
infall of the material in the inner region, i.e. below the 
Keplerian  radius $R_{K}$, and by outflow in the 
outer region above this Keplerian radius.
Figure 3 illustrates the irregular form of the dense 
compression at an arbitrarily chosen later time
(390 ksec from the initial start).
The arrows emphasize the flow divergence of the dense material both downward 
and upward from the Keplerian radius.
This evolution is most vividly illustrated through animations, which can be 
viewed on the web at:  

\noindent
{\tt 
www.bartol.udel.edu/$\sim$owocki/animations/den4wp5eta10.avi}

We have carried out similar MHD simulations for a moderately extensive 
set of combinations for the rotation and magnetic confinement 
parameters.
In all cases we find that any equatorial compressions are 
dominated by radial inflows and/or outflows, with no apparent tendency 
to form a steady, Keplerian disk.
In weak magnetic spin-up cases with $R_{A} < R_{K}$,
material trapped on closed loops generally falls back on the star, 
much as in the non-rotating models discussed in the talk by A. 
ud-Doula (see also ud-Doula 2002 and ud-Doula and Owocki 2002).
The intermediate magnetic spin-up cases with $R_{K} < R_{A} < R_{E}$ (lying 
in the  gray area of figure 1) generally show a combination of infall and 
outflow, much as described above.
Finally, for the strong magnetic spin-up models with $R_{A} > R_{E}$ 
the material density tends to build up at the tops the highest loops 
before breaking out in semi-regular epsiodes of discrete ``mass 
ejections''.

Unfortunately technical issues make it difficult to carry out 
simulations with very high magnetic fields and/or very rapid stellar rotation.
For rotations near the critical speed, the star becomes distorted into 
an oblate spheroid, thus making it difficult to formulate a static 
atmosphere lower boundary condition within our spherically symmetric 
version of the ZEUS MHD code.
For very strong magnetic fields (i.e. $\eta_{\ast} > 100$), the large 
associated Alfven speed implies a small Courant time-step within the 
explicit time-stepping of the Zeus MHD code, and so such cases become 
very computationally expensive to evolve through the required few 
characteristic flow evolution times.
Further work is thus needed to explore such cases of near-critical rotation 
and/or very strong magnetic confinement.

\section{Magnetically Confined ``Rigid-Body'' Disks}

In lieu of detailed simulations, it is however possible to infer some 
likely attributes of magnetically torqued outflows in the strong 
field limit, for which the magnetic lines-of-force can be idealized as ``rigid 
pipes'' that channel the wind outflow toward a collision near fixed 
loop tops.
For the case of a rotation aligned dipole, this effectively
concentrates material at the combined magnetic/rotational equator.
Below the Keplerian radius, such material will again tend to fall back 
to the star, but above this radius, it will be centrifugally pushed against the 
loop top, gradually building in mass until it finally has a 
sufficient rotational energy to break open the field.
By comparing the gravitational-rotational energy of this building disk to the 
magnetic field energy, one can derive an estimate for this ``break-out'' time. 
For typical stellar parameters, the ratio of this 
to a characteristic  wind flow time $\tau_{flow} = 
R_{\ast}/V_{\infty}$ scales roughly as
\begin{equation}
    { \tau_{break} \over \tau_{flow}}  \approx {
    0.01 \eta_{\ast} 
    W^{-4/3} 
    \over (r/R_{K})^{4} - (r/R_{K}) } \, .
\label{tbreakdef}
\end{equation}
Such a disk would thus have a roughly constant inner edge at the 
Keplerian radius $R_{K}$, but a highly variable outer edge.
Indeed, having any sort of disk of accumulated material that extends 
much beyond $R_{K}$ requires that the numerator in eqn. (6)
exceed unity.
In that case, we can estimate a maximum outer cutoff radius to be where 
$\tau_{break} \approx \tau_{flow}$, which solves to
\begin{equation}
    R_{out,max} \approx  { R_{A} \over \sqrt{10} \, W } \, .
\label{routmaxdef}
\end{equation}

A key point here, however, is that such disks would be quite distinct 
from a Keplerian disk, both physically and in terms of likely  obervational 
signatures.  
In particular, they seem likely to be much more variable, 
characterized by intervals with substantial radial flow speeds in the 
outer disk.
Perhaps even more notably, their rotation would follow a {\it rigid-body} law,
$v_{\phi} \sim r$, instead of the Keplerian form, $v_{\phi} \sim 1/\sqrt{r}$.

In terms of a proposed application to Be-star disks, both properties 
seem to run counter to traditional interpretations of Be-disk emission
line-profiles  (Hanuschik 1995), which in many Be stars exhibit features 
(e.g. ``central-quasi emissions''; Rivinius et al. 1999) that suggest a 
quite low upper  limit (ca. 10-20~km/s) to radial outflow speeds, and 
moreover seem  generally consistent with an azimuthal speed that 
follows a Keplerian law. 
The problems becomes particularly acute for extended rigid-body 
disks, since these imply more rapid rotation speeds than are 
typically inferred from Be emission-line-widths.
For disks with a limited radial extent, the differences from Keplerian 
rotation are less dramatic, and as explored in the poster contribution by 
D. Telfer et al., the resulting profiles could indeed be consistent with 
existing observational analyses.

However, the long-term V/R variations seen in a substantial fraction 
of Be stars do still seem best explained by a Keplerian disk 
undergoing a precession of elliptical orbits that characterize an
one-arm disk oscillation
(Savonije and Heemskerk 1993; Telting et al. 1994; Savonije 1998).
In a rigid-body disk in which the individual fluid elements of the 
disk are tied to the rotation period of the star, it is difficult to 
see how any processes could reproduce the year-long timescales of V/R 
asymmetries that require an associated long-term distinction in the 
physical properties of emitting material in a specific fixed-frame direction. 

On the other hand, such rigid-body magnetic disks may represent a good 
paradigm for the observed circumstellar emission from Bp stars 
(see, e.g. talk write-up by Groote and references therein).
The observed ca. 10~kG fields of these stars combined with their 
generally modest mass loss rates ${\dot M} \sim 10^{-9} M_{\odot}$/yr 
imply a  huge magnetic confinement parameter, of order 
$\eta_{\ast} \sim 10^{6}$.
With the associated Alfven radii of $R_{A} \approx 30 R_{\ast}$, even the 
generally modest rotations of stars with, say,  $W \approx 0.1$, 
would allow centrifugal disk support above
$R_{K} \approx 10^{2/3}  R_{\ast} \approx 4.6 R_{\ast}$,
extending perhaps out to $R_{out,max} \approx 100 R_{\ast}$.
Such low-density winds are moreover subject to ion runaway instability 
(Krticka and Kubat 2000, 2001; Owocki and Puls 2002), and this likely plays 
an important role in the abundance variations often associated with the 
circumstellar emission from Bp stars.
Future efforts should thus explore further such a wind-fed ``magnetically rigid 
disk''  (MRD) paradigm for Bp stars.

\acknowledgements

The research herein was supported in part by NASA grant NAG5-3530 and
NSF grant AST-0097983 to the University of Delaware, and by an UK-PPARC 
Visitor Fellowship to SPO.
SPO is grateful to J. Brown at University of Glasgow and A. Willis at 
University College London for their hopitality during his sabbatical visit 
to these institutions.



\begin{references}

\reference Cassinelli,ÊJ.ÊP., Brown,ÊJ.ÊC., Maheswaran,ÊM., 
           Miller,ÊN.ÊA., and Telfer,ÊD.ÊC. 2002, \apj, 578, 951.
\reference Castor, J.~I., Abbott, D.~C., \& Klein, R.~I.\ 1975, \apj, 195, 
           157 (CAK).
\reference Hanushik, R. W. 1995, \aap, 295, 423.
\reference Krti{\v c}ka, J.~\& Kub{\' a}t, J.\ 2000, \aap, 359, 983.
\reference Krti{\v c}ka, J.~\& Kub{\' a}t, J.\ 2001, \aap, 369, 222.
\reference Owocki, S. P. and Puls, J. 2002, \apj, 568, 965.
\reference Owocki, S. P. 2003, in ``Stellar Rotation'', IAU Symposium 
            215, A. Maeder and P. Eenens, eds., in press.
\reference Rivinius, T., Stefl, S., and Baade, D. 1999,  \aap, 348, 831.
\reference Savonije, ÊG.ÊJ. 1998, in {\it Cyclical Variability in Stellar Winds},
           L. Kapers and A. Fullerton, eds., Springer: Berlin, p. 337.
\reference Savonije,ÊG.ÊJ., and Heemskerk,ÊM.ÊH.ÊM. 1993, \aap, 276, 409.
\reference Telting,ÊJ.ÊH., Heemskerk,ÊM.ÊH.ÊM., Henrichs,ÊH.ÊF., 
           and Savonije,ÊG.ÊJ. 1994, \aap, 288, 558.
\reference ud-Doula, A. 2002, Ph. D. Thesis, University of Delaware.
\reference ud-Doula, A. and Owocki, S. P. 2002, \apj, 576, 413.

\end{references}
\end{document}